\documentclass[10pts,showpacs,preprint,aps]{revtex4}
\usepackage{graphicx}
\usepackage{dcolumn}
\usepackage{bm}
\usepackage{amssymb}
\usepackage{amsmath}
\begin{document}
\setcounter{page}{1}
\vskip 2cm

\title 
{Can a non-local model of gravity reproduce Dark Matter effects in agreement with MOND?}
\author
{Ivan Arraut$^{(1,2)}$}
\affiliation{$^1$Department of Physics, Osaka University, Toyonaka, Osaka 560-0043, Japan}
\affiliation{$^2$Theory Center, Institute of Particle and Nuclear Studies, KEK Tsukuba, Ibaraki, 305-0801, Japan}

\begin{abstract}
I analyze the possibility of reproducing MONDian Dark Matter effects by using a non-local model of gravity. The model was used before in order to recreate screening effects for the Cosmological Constant ($\Lambda$) value. Although the model in the weak-field approximation (in static coordinates) can reproduce the field equations in agreement with the AQUAL Lagrangian, the solutions are scale dependent and cannot reproduce the same dynamics in agreement with MOND.      
\end{abstract}
\preprint{KEK-TH-1673} 

\maketitle 
\section{Introduction}
One of the biggest problem in Cosmology is to explain the observed flatness for the galaxy rotation curves and the related observed gravitational lenses effect \cite{RC1,RC2,RC3,RC4,RC5}. Many attempts have been done in order to solve the problem, including Modified Newtonian Dynamics (MOND) \cite{1,RC6,RC7,RC8,RC9}, Modified Gravity (MOG) \cite{RC10,RC11,RC12}, non-localities \cite{Mashoon,RC13,RC14}, etc. Another big problem in Cosmology is to explain the small observed value of the Cosmological Constant $\Lambda$ or in other models, to explain the observed accelerated expansion of the universe or Dark energy \cite{DE problem,RC15}. There have been many attempts in order to explain the observed small value of $\Lambda$. Among the possibilities, we can find that the introduction of non localities can reproduce the effect of Screening of the Cosmological Constant. Some works around this issue have been proposed by S. Deser, R. P. Woodard, Odintsov and Sasaki\cite{S. Deser, MS, Odintsov}.   
If non-localities can reproduce Dark Matter effects (not necessarily MONDian) and non-localities can also screen the $\Lambda$ value, then it is natural to ask if it is possible to create a non-local model of gravity such that both effects could be incorporated. \\
If we observe that the MOND fit parameter is the $\Lambda$ scale $a_0=\frac{1}{r_\Lambda}$, with $r_\Lambda=10^{26}$ mt, then it is natural to believe that perhaps Dark Matter and Dark Energy have a common origin \cite{1,RC6,RC7,RC8,RC9}. It is natural then to suspect that the small value of $\Lambda$ is just related to the existence of Dark Matter.  
Under that philosophy, in this paper I take an already suggested non-local gravity model which introduce two scalar fields (one non-dynamical) in order to create an screening effect of the $\Lambda$ value and I then compare it with the MOND results in agreement with the AQUAL-like equation $\nabla\cdot\left(\mu\left(\frac{\vert\nabla\Phi\vert}{a_0}\right)\nabla\Phi\right)=4\pi G\rho$ \cite{Bekenstein} in order to reproduce the Dark Matter effects at least for the observed Galaxy Rotation curves. 
The AQUAL equation provides the appropriate predictions for the extragalactic phenomenology even if the AQUAL model itself cannot be realistic since it provides unphysical results \cite{URA,URA2,URA3}. It is however already known that an appropriate Relativistic version of MOND must reproduce the AQUAL equation in the weak field approximation \cite{Bekenstein}.\\ 
In this paper I find that a non-local model of gravity can reproduce the same AQUAL equations in agreement with MOND, but not the same dynamics since the interpolating function parameter $\mu$ in this model is a scale-dependent quantity because it depends on the potential $\phi$ rather than on the acceleration $\nabla\phi$. I am not concerned about the origin of the non-localities in this paper.
The paper is organized as follows: In Section \ref{eq:Lag MOND}, I make a brief review of the Lagrangian formulation of MOND. In Section \ref{eq:TMOND}, I derive the MOND field equations from the AQUAL Lagrangian. In section \ref{eq:LDB}, I analyze the behavior of the MOND field equations for large distances from the source. In section \ref{eq:RL}, I introduce the non-local model of gravity originally studied in order to create an screening effect for $\Lambda$. In section \ref{eq:NL}, I analyze the standard Newtonian limit for the Einstein's field equations with $\Lambda$ and then I analyze the Newtonian limit for the non-local model of gravity. I then find the explicit form of the interpolating function $\mu$ in terms of the potential. In section \ref{eq:56} I find the explicit solutions for $\phi$ and $\mu$ as a function of the distance $r$. I then analyze the different cases depending on the value taken by the free-parameters of the model. In section \ref{eq:comp} I make a comparison with other non-local models which were able to reproduce MOND.      

\section{Lagrangian formulation of MOND}   \label{eq:Lag MOND}

The modified dynamics assumption in agreement with Milgrom \cite{1,RC6,RC7,RC8,RC9}, can be based in the following set of minimal assumptions \cite{Bekenstein, URA,URA2,URA3}:
1). There exist a breakdown of Newtonian dynamics (second law and/or gravity) in the limit of small accelerations.\\
2). In this limit, the acceleration $\vec{a}$, of a test particle in the gravitating system is given by $\vec{a}\left(\frac{\vec{a}}{a_0}\right)\approx \vec{g}_N$, where $\vec{g}_N$ is the conventional gravitational field and $a_0$ is a constant with dimensions of acceleration.\\
3). The transition from the Newtonian regime to the small acceleration asymptotic region occurs within a range of order $a_0$ about $a_0$. The value of $a_0$ is of the same order of magnitude of $c H_0$.
The original results obtained by Milgrom \cite{1,RC6,RC7,RC8,RC9} can be described either of the following ways. A modification of the inertia:

\begin{equation}   \label{eq:1000}
m\mu\left(\frac{a}{a_0}\right)\vec{a}=\vec{F}
\end{equation}

Where $\vec{F}$ is an arbitrary static force assumed to depend on its sources in the conventional way, m is the gravitational mass of the accelerated particle and $\mu$ is the interpolating function which will be defined later. In the case of gravity, $\vec{F}=m\vec{g}_N$, where $\vec{g}_N=-\nabla\phi_N$ and $\phi_N$ is the gravitational potential deduced in the usual way from the Poisson equation.
Alternatively, MOND in agreement with \cite{1,RC6,RC7,RC8,RC9} can be described as a modification of gravity leaving the law of inertia ($m\vec{a}=\vec{F}$) intact. Then, $\vec{F}=m\vec{g}$ and $\vec{g}$ is a modified gravitational field derived from $\vec{g}_N$ using the relation:

\begin{equation}   \label{eq:1001}
\mu(g/a_0)\vec{g}=\vec{g}_N
\end{equation} 

If only gravitational forces were present, both formulations of MOND given by eqns. ($\ref{eq:1000}$) and ($\ref{eq:1001}$) would be just equivalent. However, if we consider any force in general, then the previous formulations are not the same at all. The interpolating function $\mu(x)$ satisfies the following conditions:

\begin{equation}   \label{eq:1002}
\mu(x)\approx 1\;\;\;if\;\; x>>1\;\;\;\;\;\;\;\mu(x)\approx x\;\;\; if\;\; x<<1
\end{equation}
   
Where the left side corresponds to the standard Newtonian limit. The function $\mu(x)$, can be defined in different ways. Here I will not be concerned with its definition but on its asymptotic behavior.\\
In cases of high symmetry (spherical, plane, or cylindrical), the gravitational field $\vec{g}$ as given by equation (\ref{eq:1001}) is derivable from a scalar potential $\phi$. However, in the most general cases this is not possible. As has been already explained by Milgrom in his paper \cite{Milgrom1}, MOND cannot be considered as a theory, but only a successful phenomenological scheme for which an underlying theory can be constructed. One of the reasons is for example, that inside MOND theory there is no momentum conservation. In fact, momentum is only conserved approximately as far as the mass of the test body is much smaller than the source one \cite{URA,URA2,URA3}. \\
Milgrom and Bekenstein have already derived a Lagrangian formulation for MOND, where $\mu$ and $a_0$ are introduced by hand.
One of the purposes for constructing a fundamental theory which can contain MOND as a non-relativistic limit is to obtain the interpolating function $\mu(x)$ in terms of some fundamental quantities. This is one of the motivations for this manuscript. A Lagrangian formulation for MOND solves the momentum conservation problem associated typically to the original MOND version. A Lagrangian formulation also enables to calculate the dynamics of an arbitrary non-relativistic system \cite{URA,URA2,URA3}.\\
There are two important assumptions inside the MOND theory, they are:\\
1). A composite particle (star or a cluster of stars) moving in an external field, say a galaxy, moves like a test particle according to MOND. Even if within the body, the relativistic accelerations are large. This assumption is possible as far as the mass of the test particle is much smaller than the mass of the galaxy.\\
2). When a system is accelerated as a whole in an external field, the internal dynamics of the system is affected by the presence of an external field (even when this field is constant without tidal forces). In particular, in the limit when the external (center of mass) acceleration of the system becomes much larger than the MOND scale $a_0$, the internal dynamics approaches to the Newtonian behavior even when the accelerations within the system are much smaller than $a_0$.\\
This second observation due to the Milgrom proposal is quite interesting and one of the motivations for this manuscript since it seems that the internal dynamics of the system can in principle be affected by the presence of some external field. This could in principle be done due to a non-local connection between the internal and external dynamics.\\
Normally Newtonian gravity is recovered at the non-relativistic regime of General Relativity (GR), and of a number of other relativistic theories of gravity. It is however, necessary to construct a new relativistic version of GR such that MOND could be recovered as a natural non-relativistic limit \cite{URA,URA2,URA3}. This new version could be constructed by considering two possibilities. 1). Additional degrees of freedom. 2). Non-localities. If we choose to explain the origin of Dark Matter by introducing non-localities, we must be able to explain the origin of such effects. In this manuscript I will not be concerned with the origin of non-localities, I will introduce them in an arbitrary way and then we will write the MOND interpolating function in terms of the field generating the non-local effects.\\
There are two very important reasons to construct a Relativistic version for MOND. 1). To help incorporate principles of MOND into the framework of modern theoretical physics. 2). To provide tools for investigating cosmology in light of MOND. \cite{URA,URA2,URA3}

\section{The MOND field equations}   \label{eq:TMOND}

In Newtonian gravity test bodies move with an acceleration equal to $\vec{g}_N=-\nabla\phi_N$, where $\phi_N$ is the Newtonian gravitational potential. It is determined by the Poisson equation $\nabla^2\phi_N=4\pi G\rho$, where $\rho$ is the mass density which produces $\phi_N$. The Poisson equation may be derived from the Lagrangian:

\begin{equation}   \label{eq:1003}
L_N=-\int d^3r(\rho\phi_N+(8\pi G)^{-1}(\nabla\phi_N)^2)
\end{equation}  

Milgrom and Bekenstein suggested that in searching for a modification of this theory, we will want to retain the notion of a single potential $\phi_N$ from which acceleration derives. We want $\phi_N$ to be arbitrary up to an arbitrary additive constant. The most general modification of $L_N$ which yield these features is:

\begin{equation}   \label{eq:1004}
L=-\int d^3r\left(\rho\phi+(8\pi G)^{-1}a_0^2f\left(\frac{(\nabla\phi)^2}{a_0^2}\right)\right)
\end{equation}

Where $f(x^2)$ is an arbitrary function. The scale of acceleration is necessary unless we are in the Newtonian case. If we perform the variation of L with respect to $\phi$, with variation of $\phi$ vanishing on the boundaries, we get:

\begin{equation}   \label{eq:1005}
\vec{\nabla}\cdot\left(\mu\left(\frac{\vert\vec{\nabla}\phi\vert}{a_0}\right)\vec{\nabla}\phi\right)=4\pi G\rho
\end{equation}

Where $\mu(x)=f'(x^2)$. Eq. (\ref{eq:1005}) is the equation determining the modified potential. A test particle is assumed to have acceleration $\vec{g}=-\vec{\nabla}\phi$. We supplement equation (\ref{eq:1005}) by the boundary condition $\vert\vec{\nabla}\phi\vert\to0$ as $r\to\infty$. \\
It is useful to write the field equation in terms of the modified Newtonian field $\vec{g}_N=-\vec{\nabla}\phi_N$, for the same mass distribution, which satisfies the Poisson equation. By eliminating $\rho$ in equation (\ref{eq:1005}), we get:

\begin{equation}   \label{eq:1006}
\vec{\nabla}\cdot\left(\mu\left(\frac{\vert\vec{\nabla}\phi\vert}{a_0}\right)\vec{\nabla}\phi-\vec{\nabla}\phi_N\right)=0
\end{equation}

The expression in parenthesis, must then be a curl, since its divergence is zero. Then we can write:

\begin{equation}   \label{eq:1007}
\mu\left(\frac{g}{a_0}\right)\vec{g}=\vec{g}_N+\vec{\nabla}\times\vec{h}
\end{equation}

It has already been demonstrated by Bekenstein and Milgrom that the present theory satisfies the basic assumptions of MOND and that the curl term vanishes exactly for the spherically symmetric case and it vanishes at least as fast as $\frac{1}{r^3}$ at large distances from the source in general cases. In the next section I will consider the review of this point.

\section{The field equations at large distances from the source}   \label{eq:LDB}   

In agreement with \cite{URA,URA2,URA3}, let's consider a bound density distribution of total mass M with the origin at the center of mass. Following the Bekenstein and Milgrom notation, let's define the vector field $\vec{u}$ as $\vec{u}\equiv\vec{\nabla}\phi_N-\mu\left(\frac{\vert\vec{\nabla}\phi\vert}{a_0}\right)\vec{\nabla}\phi$. For the reasons explained in the previous section, the vector $\vec{u}$, satisfies $\vec{\nabla}\cdot\vec{u}=0$ and it vanishes at infinity. It is possible then to write $\vec{u}$ in terms of the vector potential $\vec{A}$:

\begin{equation}   \label{eq:1008}
\vec{u}=\vec{\nabla}\times\vec{A}\;\;\;\;\;\;\;\;\vec{A}(\vec{r})=(4\pi)^{-1}\int\frac{\vec{\nabla}'\times \vec{u}(\vec{r'})}{\vert\vec{r}-\vec{r'}\vert}d^3r'
\end{equation}   
   
The only term with an $r^{-2}$ behavior at infinity which $\vec{u}$ can have is $\vec{u}^{(2)}=\vec{\nabla}\times(r^{-1}\vec{B})=-r^{-3}\vec{r}\times\vec{B}$. Where $\vec{B}=(4\pi)^{-1}\int \vec{\nabla}'\times\vec{u}'d^3r'$, which is the lowest order in the multipole expansion (\ref{eq:1008}). \\
If we can demonstrate that $\vec{B}=0$, then we get that the lowest contributing multiple term to $\vec{u}$ vanishes at least as fast as $r^{-3}$. In the limit of large r, we have:

\begin{equation}   \label{eq:1009}
\mu\left(\frac{\vert\vec{\nabla}\phi\vert}{a_0}\right)\vec{\nabla}\phi=\vec{\nabla}\phi_N-\vec{u}=r^{-3}(GM\vec{r}+\vec{r}\times\vec{B})+\vec{O}(r^{-3})
\end{equation}

Taking the absolute value, we get:

\begin{equation}   \label{eq:1010}
\mu\left(\frac{\vert\vec{\nabla}\phi\vert}{a_0}\right)\vert\vec{\nabla}\phi\vert=(\vert\vec{\nabla}\phi_N\vert^2+\vert\vec{u}\vert^2)^{1/2}
\end{equation}

This expression can be translated into:

\begin{equation}   \label{eq:1011}
\mu\left(\frac{\vert\vec{\nabla}\phi\vert}{a_0}\right)\vert\vec{\nabla}\phi\vert=\left(\left(\frac{GM}{r^2}\right)^2+\frac{B^2sin^2}{r^4}\theta\right)^{1/2}
\end{equation}

As $r\to\infty$, the full MONDian regime operates and we can assume that $\mu\left(\frac{\vert\vec{\nabla}\phi\vert}{a_0}\right)=\frac{\vert\vec{\nabla}\phi\vert}{a_0}$. In such a case, the expression (\ref{eq:1011}) becomes:

\begin{equation}   \label{eq:1012}
\vert\vec{\nabla}\phi\vert=\frac{a_0^{1/2}}{r}\left(G^2M^2+B^2sin^2\theta\right)^{1/4}
\end{equation}

Assuming again the MONDian regime and replacing the previous expression inside of (\ref{eq:1009}), we get:

\begin{equation}   \label{eq:1013}
\vec{\nabla}\phi=a_0^{1/2}r^{-2}\frac{(GM\vec{r}+\vec{r}\times\vec{B})}{(G^2M^2+B^2sin^2\theta)^{1/4}}+\vec{O}(r^{-2})
\end{equation}

Here $\theta$ is the angle between $\vec{r}$ and $\vec{B}$ which we can take along the z-axis without lost of generality. Requiring now that the azimuthal component of $\vec{\nabla}\times(\vec{\nabla}\phi)$ vanishes, gives $\vec{B}=0$ \cite{URA,URA2,URA3}. This means that $\vec{u}$ vanishes at large distances from a mass, at least as $\vec{O}(r^{-3})$. With this result, the equation (\ref{eq:1007}) as $r\to\infty$, becomes:

\begin{equation}   \label{eq:1014}
\mu\left(\frac{g}{a_0}\right)\vec{g}=\vec{g}_N+\vec{O}(r^{-3})
\end{equation} 

Consistent with the MOND predictions explained in eqns. (\ref{eq:1000}) and (\ref{eq:1001}). As $r\to\infty$, we get:

\begin{equation}   \label{eq:1015}
\vec{g}\to-\frac{(GM a_0)^{1/2}}{r^2}\vec{r}+\vec{O}(r^{-2})
\end{equation}

In this limit, the potential becomes:

\begin{equation}   \label{eq:1016}
\phi\to(GMa_0)^{1/2}ln\left(\frac{r}{r_0}\right)+O(r^{-1})
\end{equation}

Where $r_0$ is an arbitrary radius. This potential leads to an asymptotically constant circular velocity $V_\infty=(GMa_0)^{1/4}$ as it is observed in the outskirts of spiral galaxies.\\
The field equation (\ref{eq:1005}) is nonlinear and difficult to solve in general. However, in cases of high symmetry, the curl term in equation (\ref{eq:1007}) vanishes identically and we have the exact result $\mu\left(\frac{g}{a_0}\right)\vec{g}=\vec{g}_N$ which is identical to equation (\ref{eq:1001}). For systems for high degree of symmetry, then the solution for $\phi$ is straightforward and all the results obtained from the standard Newtonian theory can then be extended to the present formalism.\\
For example, the acceleration field at a distance r from the center in a spherical system depends only on the total mass $M(r)$, interior to r (in agreement with the Gauss' theorem), and in fact is given by $\mu\left(\frac{g}{a_0}\right)\vec{g}=-\frac{M(r)G\vec{r}}{r^3}$.\\
The field equation (\ref{eq:1005}) is analogous to the equation for the electrostatic potential in a nonlinear isotropic medium in which the dielectric coefficient is a function of the electric field strength \cite{URA,URA2,URA3}.\\
The field equation (\ref{eq:1005}) is also equivalent to the stationary flow equations of an irrotational fluid which has a density $\hat{\rho}=\mu\left(\frac{\vert\vec{\nabla}\phi\vert}{a_0}\right)$, a negative pressure $\hat{P}=-\frac{1}{2}a_0^2f\left(\frac{(\vec{\nabla}\phi)^2}{a_0^2}\right)$, flow velocity $\hat{\vec{v}}=\vec{\nabla}\phi$, and a source distribution $\hat{S}(\vec{r})=4\pi G\rho$. The fluid satisfies an equation of state $\hat{P}(\hat{\rho})=-\frac{1}{2}a_0^2f\left([(\mu^{-1}(\hat{\rho})]^2\right)$. \\
An equation of the same form as equation (\ref{eq:1005}) has been studied to describe classical models of quark confinement using a very different form of the function $\mu$ at both, large and small values of its argument \cite{Piran}.
The conservation laws and other results related to the Lagrangian formulation of MOND can be found in \cite{URA,URA2,URA3}. In this manuscript I will omit such analysis.

\section{A Non-local model for gravity}   \label{eq:RL}

The non-local action suggested in \cite{MS} is given by:

\begin{equation}   \label{eq:1}
S=\int d^4x\sqrt{-g}\left(\frac{1}{2\kappa^2}(R(1+f(\square^{-1}R))-2\Lambda)+l_{matter}(Q, g)\right)
\end{equation}

Where f is some function, $\square$ is just the D'Alembertian for the scalar field, $\Lambda$ is the Cosmological Constant which is supposed to be screened by the introduced non-localities and Q corresponds to the matter fields. We can rewrite the action by introducing two scalar fields $\psi$ and $\zeta$ as follows \cite{MS}:

\begin{eqnarray}   \label{eq:2}
S=\int d^4x\sqrt{-g}\left(\frac{1}{2\kappa^2}(R(1+f(\psi))-\zeta(\square\psi-R)-2\Lambda)+l_{matter}\right)\nonumber \\
=\int d^4x\sqrt{-g}\left(\frac{1}{2\kappa^2}(R(1+f(\psi)+\zeta)+g^{\mu \nu}\partial_\mu\zeta\partial_\nu\psi-2\Lambda)+l_{matter}\right)
\end{eqnarray}

If we vary the above action with respect to $\zeta$, then $\square\psi=R$ or $\psi=\square^{-1}R$.  The variation with respect to the metric is:

\begin{eqnarray}   \label{eq:3}
0=\frac{1}{2}g_{\mu \nu}\left(R(1+f(\psi)+\zeta)+g^{\alpha \beta}\partial_\alpha\zeta\partial_\beta\psi-2\Lambda\right)-R_{\mu \nu}(1+f(\psi)+\zeta)\nonumber\\
-\frac{1}{2}(\partial_\mu\zeta\partial_\nu\psi+\partial_\mu\psi\partial_\nu\zeta)-(g_{\mu \nu}\square-\nabla_\mu\nabla_\nu)(f(\psi)+\zeta)+\kappa^2T_{\mu \nu}
\end{eqnarray}

And the variation with respect to $\psi$ gives:

\begin{equation}   \label{eq:4}
0=\square\zeta-f'(\psi)R
\end{equation}

The explicit solutions for the previous equations, can be found if we introduce a metric. In this manuscript, I will focus on spherical symmetric solutions. I will introduce our metric in the next section. 

\subsection{The ghost free condition}   \label{eq:GFC}

In \cite{MS, MS2}, it was found that after a conformal transformation to the Einstein frame, we get:

\begin{equation}   \label{eq:25}
\tilde{g}_{\mu \nu}=\Omega^2 g_{\mu \nu}\;\;\;\;\;\tilde{R}=\frac{1}{\Omega^2}(R-6(\square ln\Omega+g^{\mu \nu}\nabla_\mu ln\Omega\nabla_\nu ln\Omega))
\end{equation} 
 
\begin{equation}   \label{eq:26}
\Omega^2=\frac{1}{1+f(\psi)+\zeta}
\end{equation} 

Which gives an action given by \cite{MS}:

\begin{equation}   \label{eq:27}
S=\int d^4x\sqrt{-g}\left(\frac{1}{2\kappa^2}(\hat{R}-6g^{\mu \nu}\nabla_\mu \phi' \nabla_\nu \phi'+e^{2\phi'}g^{\mu \nu}\nabla_\mu\zeta\nabla_\nu\psi-2e^{4\phi'}\Lambda)+e^{4\phi'}l_{matter}(Q;e^{2\phi'}g)\right)
\end{equation}

Where:

\begin{equation}   \label{eq:28}
\phi'\equiv ln\Omega=-\frac{1}{2}ln(1+f(\psi)+\zeta)
\end{equation}

and $\hat{R}$ is the resulting Ricci scalar after performing the transformation (\ref{eq:25}). The condition for gravity to have a normal sign is:

\begin{equation}   \label{eq:29}
1+f(\psi)+\zeta>0
\end{equation}

If $\phi'$ and $\psi$ are considered to be the independent fields, then:

\begin{equation}   \label{eq:30}
\zeta=e^{-2\phi'}-(1+f(\psi))
\end{equation}

Then, in terms of the new set of independent variables, the action is:

\begin{eqnarray}   \label{eq:31}
S=\int\sqrt{-g}\frac{1}{2\kappa^2}(R-6\nabla^\mu\phi'\nabla_\mu\phi'-2\nabla^\mu\phi'\nabla_\mu\psi-e^{2\phi'}f'(\psi)\nabla^\mu \psi \nabla_\mu \psi-2e^{4\phi'}\Lambda)\nonumber\\
+e^{4\phi'}l_{matter}(Q;e^{2\phi'g})
\end{eqnarray}

The ghost free condition is simply:

\begin{equation}   \label{eq:32}
f'(\psi)>\frac{1+f(\psi)+\zeta}{6}>0
\end{equation}

\section{The Newtonian limit in the standard case in S-dS metric}   \label{eq:NL} 

In \cite{MS}, the metric is assumed to be FLRW. In this case, as we are concerned with the Dark Matter effects and we want to compare with the MONDian case, then I will assume that the space time metric corresponds to the Newton-Hooke space, which is just the Newtonian limit for the Schwarzschild-de Sitter space. Explicitly in eqs. (\ref{eq:3}) and (\ref{eq:4}), I will assume the metric to be:

\begin{equation}   \label{eq:5}
ds^2=-\left(1-\frac{2GM}{r}-\frac{1}{3}\frac{r^2}{r_\Lambda^2}\right)dt^2+\left(1-\frac{2GM}{r}-\frac{1}{3}\frac{r^2}{r_\Lambda^2}\right)^{-1}dr^2+r^2d\Omega^2
\end{equation}

With $d\Omega^2=d\theta^2+sin^2\theta d\phi^2$. I will work under the condition $r_s<<r<<r_\Lambda$. Under that condition, the weak field approximation is justified. Under the Weak Field approximation, we have to satisfy the standard results:   

\begin{equation}   \label{eq:6}
G_{00}^{(1)}\approx \square g_{00}\approx R^{(1)}=2R_{00}^{(1)}\;\;\;
\end{equation}

Where the weak field approximation for the Ricci tensor is given by:

\begin{equation}   \label{eq:77}
R_{\mu \nu}^{(1)}\equiv \frac{1}{2}(\square h_{\mu \nu}-\partial^\lambda \partial_\mu h_{\lambda \nu}-\partial^\lambda \partial_\nu h_{\lambda \mu}+\partial_\mu \partial_\nu h)  
\end{equation}

And then, the first order Einstein's equations become:

\begin{equation}   \label{eq:777}
R_{\mu \nu}^{(1)}-\frac{1}{2}\eta_{\mu \nu}R^{(1)}+\eta_{\mu \nu}\Lambda=-8\pi G_N T_{\mu \nu}^{(1)}
\end{equation}

With the metric given by (\ref{eq:5}), then we get:

\begin{equation}   \label{eq:7}
\nabla^2g_{00}=-8\pi G\rho+2\Lambda
\end{equation}

If $g_{\mu \nu}\approx \eta_{\mu \nu}+h_{\mu \nu}$, then $\nabla g_{00}\approx \nabla h_{00}$. Thus:

\begin{equation}   \label{eq:8}
\nabla^2\phi=4\pi G_N\rho-\Lambda
\end{equation}

With $h_{00}=-2\phi=h_{ij}$ and $T_{00}\approx \rho$. The spherical symmetry of the metric (\ref{eq:5}) is important since it implies that the curl term in equation (\ref{eq:1007}) can be ignored in agreement with the analysis performed in the previous section. The results of this section will be used in the field equations \ref{eq:3} and \ref{eq:4}. 

\subsection{The weak field approximation in non-local gravity and its relation with MOND}   \label{eq:WFAMOND}

In agreement with Bekenstein \cite{Bekenstein}, we have to satisfy at the Newtonian limit an equation similar to the AQUAL given already in equation (\ref{eq:1005}). We rewrite it as follows:

\begin{equation}   \label{eq:13}
\nabla^2\phi\approx \mu^{-1}(\kappa^2\rho-\Lambda)-\mu^{-1}(\nabla\phi).\nabla\left(\mu\left(\frac{\vert\nabla\phi\vert}{a_0}\right)\right)
\end{equation}  

For the Newtonian limit of the field equations (\ref{eq:3}), I will make the expansions up to second order in the potential $\phi$. Even if the second order terms are most likely negligible, I will keep them in order to get a more accurate result. Take into account that the non-localities, represented by $f(\psi)+\zeta$ in eq. (\ref{eq:3}) reproduce an amplification of the non-linearities related to the space time curvature and it includes the second order contributions. This amplification will however depend on a parameter $\gamma$ which will be defined later. The 0-0 component of eq. (\ref{eq:3}) is then given by:

\begin{eqnarray}   \label{eq:14}
0\approx-\frac{1}{2}\left(R^{(1)}(1+f(\psi)+\zeta)+\nabla_r\zeta\nabla_r\psi-2\Lambda\right)-\phi R^{(1)}(1+f(\psi)+\zeta)\nonumber\\
-R_{00}^{(1)}(1+f(\psi)+\zeta)
+(1+2\phi)\square (f(\psi)+\zeta)+\nabla_0\nabla_0(f(\psi)+\zeta)-\kappa^2\rho
\end{eqnarray}

Where we have used $T_{00}=-\rho$. In this approach, we neglect the time-dependence of the scalar fields. However we take into account the curvature effects through the Christoffel connections. We can write $\zeta$ in terms of $f(\psi)$ if we solve the equation (\ref{eq:4}). For that purpose, we assume an exponential solution like $f(\psi)=f_0 e^{\gamma\psi}$ as has been suggested in \cite{MS}. Then the following relations are true:

\begin{equation}   \label{eq:18}
\nabla_\mu f(\psi)=\gamma f(\psi)\nabla_\mu\psi\;\;\;\;\;\square f(\psi)=\gamma f(\psi)\square\psi+\gamma^2f(\psi)\nabla_\mu\psi\nabla^\mu\psi
\end{equation}     

We can then prove that the solution for eq. (\ref{eq:4}) if we expand both sides of the equation and then compare the same order of magnitude terms. The resulting equation is:

\begin{equation}   \label{eq:Corr}
2\nabla\phi\cdot\nabla\zeta+\frac{2}{r}\frac{\partial\zeta}{\partial r}+\frac{\partial^2\zeta}{\partial r^2}\approx-\gamma f(-2\phi)\left(4\left(\frac{\partial\phi}{\partial r}\right)^2+\frac{4}{r}\frac{\partial\phi}{\partial r}+2\frac{\partial^2\phi}{\partial r^2}\right)
\end{equation}

The solution for $\zeta$ is (ignoring second order contributions):

\begin{equation}   \label{eq:19}
\zeta(\psi)\approx f(\psi)
\end{equation}

Where we have used the Lagrange multiplier condition $R=\square\psi$. If we replace the solution for $\zeta(\psi)$, taking into account eq. (\ref{eq:4}) and the Lagrange multiplier condition, eq. (\ref{eq:14}) becomes: 

\begin{eqnarray}   \label{eq:20}
0\approx -\frac{1}{2}\left(R^{(1)}(1+2f(\psi))+\gamma f(\psi)(\nabla\psi)^2-2\Lambda\right)-\phi R^{(1)}(1+2f(\psi))-R_{00}^{(1)}(1+2f(\psi))\nonumber\\
+2\gamma(1+2\phi)f(\psi)\square\psi+2\nabla_0\nabla_0f(\psi)-\kappa^2\rho
\end{eqnarray}

The Christoffel connection component is given by $\Gamma^r_{00}\approx \partial_r\phi=\nabla_r\phi$. I will take the spatial components of the Einstein's equations as given by the standard Newtonian approach as it is explained in the standard textbooks. In such a case, I will take the Ricci tensor and the curvature scalars as:

\begin{equation}   \label{eq:15}
R^{(1)}=2R^{(1)}_{00}\approx \square g_{00}=\square h_{00}=-2\square\phi\approx -4(\nabla\phi)\cdot(\nabla\phi)-2\nabla^2_M\phi-4\phi\nabla^2_M\phi
\end{equation}

Note that we are just rewriting the result (\ref{eq:77}) for the case of a static potential. Note also that in the standard Newtonian approach $\square h_{00}=\nabla^2_M h_{00}$, where the subindex $M$ makes reference to the Minkowskian case. However, in this case I consider the expansion up to second order and it includes the curvature effects obtained from the Christoffel connections. In principle, the scalar curvature and the Ricci tensor when expanded up to second order are given by:

\begin{equation}   \label{eq:16}
R\approx R^{(1)}+R^{(2)}\;\;\;\;\;R_{\mu \nu}\approx R^{(1)}_{\mu \nu}+R^{(2)}_{\mu \nu}
\end{equation}  

Up to first order, then the approximation $\psi=-2\phi$ is valid in agreement with eq. (\ref{eq:15}) and the Lagrange multiplier condition. On the other hand, eq. (\ref{eq:20}) expanded up to second order is equivalent to:

\begin{equation}   \label{eq:17}
\nabla^2_M\phi\approx \mu^{-1}(\kappa^2\rho-\Lambda)-\mu^{-1}\nabla\phi\cdot\nabla\phi(4\omega)
\end{equation}

Where $\mu(\phi)$ and $\omega$ are defined by:

\begin{equation}   \label{eq:55}
\mu(\phi)=2(1+2f(\psi)(1+3\phi-2\gamma-8\phi\gamma)+3\phi)\;\;\;\;\;\omega=1+2f(\psi)\left(1-\frac{3}{4}\gamma\right)
\end{equation}

We can observe that eq. (\ref{eq:17}) has the same form of eq. (\ref{eq:13}) which describes the MONDian dynamics. There is however a fundamental difference since in the non-local model, $\mu$ defined in eq. (\ref{eq:55}) is just a function of the potential rather than a function of the acceleration ($\vert\nabla\phi\vert$) as it is the case in the MONDian dynamics. For the weak field approximation, the following approximations for the eqs. (\ref{eq:55}) are valid:

\begin{equation}   \label{eq:This}
\mu(\phi)\approx 2(1+2f_0(1+3\phi-2\gamma-8\phi\gamma)+3\phi-4\gamma f_0\phi(1-2\gamma))\;\;\;\;\;\omega\approx 1+2f_0\left(1-\frac{3}{4}\gamma\right)
\end{equation}

\section{Explicit solutions for $\phi$ and $\mu$}   \label{eq:56}  

I will compute the explicit solutions for $\mu$ and $\phi$ in agreement with the equation (\ref{eq:17}). Then different regimes will be explored (different values for $\gamma$) and I will identify some special values for $\gamma$. It is simpler to start by solving $\mu$. For that purpose we have to find the solutions for the following equations in agreement with the result (\ref{eq:55}) for the weak field approximation:

\begin{equation}   \label{eq:57}
\nabla\mu=-4\nabla\phi\left((-3+8\gamma)f_0-\frac{3}{2}+2f_0\gamma(1-2\gamma)\right)
\end{equation}
 
And: 

\begin{equation}   \label{eq:58}
\nabla^2\mu=-4\nabla^2\phi\left((-3+8\gamma)f_0-\frac{3}{2}+2f_0\gamma(1-2\gamma)\right)
\end{equation}

Then we can write eq. (\ref{eq:17}) in terms of $\mu$. In vacuum the result is:

\begin{equation}   \label{eq:59}
\mu\nabla_M^2\mu\approx C\nabla\mu\cdot\nabla\mu
\end{equation}

Where we have defined $C$ as:

\begin{equation}   \label{eq:60}
C=\frac{\omega}{\left((-3+8\gamma)f_0-\frac{3}{2}+2f_0\gamma(1-2\gamma)\right)}
\end{equation}

The general solution for $\mu$ is given by:

\begin{equation}   \label{eq:61}
\mu(r)=A\left(\frac{-1+C}{r}+B\right)^{\frac{1}{1-C}}
\end{equation}

Note that this solution is valid for $C\neq1$. As $C=1$, eq. (\ref{eq:59}) becomes:

\begin{equation}   \label{eq:62}
\mu\nabla_M^2\mu\approx \nabla\mu\cdot\nabla\mu
\end{equation}
  
The solution for this equation is:

\begin{equation}   \label{eq:63}
\mu(r)=De^{-E/r}
\end{equation}

\subsection{Solutions for $\mu$ and $\phi$ for special values of $\gamma$}

There are different possible solutions for $\phi$ and $\mu$ in agreement with the results obtained in the previous section. In vacuum and ignoring the Cosmological Constant $\Lambda$, we can write the equation (\ref{eq:17}) as follows:

\begin{equation}   \label{eq:64}
\mu\nabla^2_M\phi=C(\nabla\phi)\cdot(\nabla\mu)
\end{equation}

Where we have used the results obtained in eq. (\ref{eq:57}) and the definition (\ref{eq:60}). I will analyze some relevant results summarized in the following table:

\noindent      
\begin{table}[h]   
\caption{\label{tab2} Relevant values for $C$ as a function of the parameter $\gamma$ and $\omega$}.  
\begin{tabular}{|l|l|l|}
\hline
\;\;C & $\;\;\;\;\;\;\;\;\;\;\;\;\;\;\;\gamma$&\;\;\;\;\;\;\;\;\;\;\;\;\;\;\;\;\;\;\;\;\;\;\;\;\;\;\;\;\;\;\;\;\;\;\;\;$\omega$ \\ 
\hline
\;\;$0$ & $\;\;\;\;\;\;\;\;\;\;\;\;\;\;\;\infty$&\;\;\;\;\;\;\;\;\;\;\;\;\;\;\;\;\;\;\;\;\;\;\;\;\;\;\;\;\;\;\;\;\;\;$-\infty$  \\ 
\hline
\;\;$0$ &$\;\;\;\;\;\;\;\;\;\;\;\;-\infty$&$\;\;\;\;\;\;\;\;\;\;\;\;\;\;\;\;\;\;\;\;\;\;\;\;\;\;\;\;\;\;\;\;\;\;\;\;\infty$\\ 
\hline
\;\;1&\;\;$\;\frac{23f_0-\sqrt{f_0}\sqrt{-160+209f_0}}{16f_0}$&\;\;\;\; $1-2f_0-\frac{3}{32}\left(23f_0-\sqrt{f_0}\sqrt{-160+209f_0}\right)$\\
\hline 
\;\;1&\;\;$\frac{23f_0+\sqrt{f_0}\sqrt{-160+209f_0}}{16f_0}$&\;\;\;\; $1-2f_0-\frac{3}{32}\left(23f_0+\sqrt{f_0}\sqrt{-160+209f_0}\right)$\\
\hline 
\;-1&\;\;$\frac{17f_0+\sqrt{f_0}\sqrt{-32+225f_0}}{16f_0}$&\;\;\;\;\; $1-2f_0-\frac{3}{32}\left(17f_0+\sqrt{f_0}\sqrt{-32+225f_0}\right)$\\
\hline
\;-1&\;\;$\frac{17f_0-\sqrt{f_0}\sqrt{-32+225f_0}}{16f_0}$&\;\;\;\; \;$1-2f_0-\frac{3}{32}\left(17f_0-\sqrt{f_0}\sqrt{-32+225f_0}\right)$\\
\hline
\;$\infty$&$\;\;\;\;\frac{5f_0+\sqrt{-6f_0+13f_0^2}}{4f_0}$&\;\;\;\; \;\;\;$1-2f_0-\frac{3}{8}\left(5f_0+\sqrt{-6f_0+13f_0^2}\right)$\\
\hline
$-\infty$&$\;\;\;\;\frac{5f_0-\sqrt{-6f_0+13f_0^2}}{4f_0}$&\;\;\;\; \;\;\;$1-2f_0-\frac{3}{8}\left(5f_0-\sqrt{-6f_0+13f_0^2}\right)$\\
\hline
\;\;0&\;\;\;\;\;\;\;\;$\;\frac{2}{3f_0}+\frac{4}{3}$&\;\;\;\;\;\;\;\;\;\;\;\;\;\;\;\;\;\;\;\;\;\;\;\;\;\;\;\;\;\;\;\;\;\;\;\;0\\
\hline
Min&$\frac{8+16f_0-\sqrt{2}\sqrt{32+35f_0-58f_0^2}}{12f_0}$&$1-2f_0-\frac{3}{24}\left(8+16f_0-\sqrt{2}\sqrt{32+35f_0-58f_0^2}\right)$\\
\hline
Max&$\frac{8+16f_0+\sqrt{2}\sqrt{32+35f_0-58f_0^2}}{12f_0}$&$1-2f_0-\frac{3}{24}\left(8+16f_0+\sqrt{2}\sqrt{32+35f_0-58f_0^2}\right)$\\
\hline
\;\;$\frac{1}{2}$&$\;\;\frac{13f_0-\sqrt{-56f_0+57f_0^2}}{8f_0}$&\;\;\;\;\; $1-2f_0-\frac{3}{16}\left(13f_0-\sqrt{-56f_0+57f_0^2}\right)$\\
\hline
\;\;$\frac{1}{2}$&$\;\;\frac{13f_0+\sqrt{-56f_0+57f_0^2}}{8f_0}$&\;\;\;\;\; $1-2f_0-\frac{3}{16}\left(13f_0+\sqrt{-56f_0+57f_0^2}\right)$\\
\hline
\end{tabular}
\end{table}     
\noindent

Where $Min$ and $Max$ correspond to a local minimum and a local maximum respectively for the parameter C as can be easily verified. If we replace the result (\ref{eq:61}) inside the definition of $\mu$ given in eq. (\ref{eq:55}), we then obtain the solution for $\phi$ consistent with eq. (\ref{eq:62}). Up to first order, the result is:

\begin{equation}   \label{eq:65}
\phi=A(\gamma)\left(\frac{-1+C}{r}+B\right)^{\frac{1}{1-C}}-2\left(\frac{1+2f_0(1-2\gamma)}{2(3+2f_0(3-10\gamma+4\gamma^2))}\right)
\end{equation}

Where $A(\gamma)$ is defined as:

\begin{equation}   \label{eq:66}
A(\gamma)=-\left(\frac{A}{4}\right)\frac{C}{\omega}\;\;\;\;\; 
\end{equation}

The same result for the case $C=1$ is:

\begin{equation}   \label{eq:67}
\phi=A(\gamma)e^{-D/r}-2\left(\frac{1+2f_0(1-2\gamma)}{2(3+2f_0(3-10\gamma+4\gamma^2))}\right)
\end{equation}

Where $D$ is just another integration constant and $\omega$ has to be evaluated for the case $C=1$. The case $C=1$ corresponds to two different values for the parameter $\gamma$ as can be seen from Table \ref{tab2}. The equations (\ref{eq:65}) and (\ref{eq:67}) can be rewritten in a compact form as:

\begin{equation}   \label{eq:0}
\phi=-\left(\frac{A}{4}\right)\left(\frac{C}{\omega}\right)\left(\frac{-1+C}{r}+B\right)^{\frac{1}{1-C}}+\frac{1}{2}\left(\frac{C}{\omega}\right)(1+2f_0(1-2\gamma))
\end{equation}

And:

\begin{equation}   \label{eq:0}
\phi=-\left(\frac{A}{4}\right)\left(\frac{1}{\omega}\right)e^{-D/r}+\frac{1}{2}\left(\frac{1}{\omega}\right)(1+2f_0(1-2\gamma))
\end{equation}

In both cases, the condition $\omega\neq0$ is satisfied. If $\omega=0$, then $C$ can take 3 different values in agreement with the Table \ref{tab2}. The standard Newtonian behavior is recovered for the case $C=0=\omega$. There will be values of $\gamma$ for which the potential $\phi$ will be attractive and other values for which it will be repulsive.  

\subsection{Special cases for different values of the parameter $\gamma$ and the ghost-free condition}

If we calculate the gradient of the potential $\phi$. Without loss of generality, we can set $B=0$ for the cases $C\neq-1$ since this constant does not affect the dynamics at large distances. If we calculate the gradient from eq. (\ref{eq:65}) as $B=0$, we get: 

\begin{equation}   \label{eq:68}
\nabla\phi=-\left(\frac{A}{4}\right)\left(\frac{C}{\omega}\right)\left(\frac{-1+C}{r}\right)^{\frac{C}{1-C}}\frac{1}{r^2}
\end{equation}

For $C\neq-1$. This potential can be attractive or repulsive depending of the value of the ratio $\frac{C}{\omega}$ and the relative sign of $C$ with respect to $-1$ for the terms inside the parenthesis. On the other hand, the gradient for the case $C=1$ is taken from eq. (\ref{eq:0}) and it is given by:

\begin{equation}   \label{eq:69}
\nabla\phi=-\left(\frac{A}{4}\right)\left(\frac{D}{\omega}\right)\left(\frac{e^{-D/r}}{r^2}\right)
\end{equation}
 
Which is attractive or repulsive depending on the sign of $\omega$. 

\subsection{The relevant cases for the potential}

If we take into account that dynamically the potential satisfies the condition $\vec{\nabla}\phi=\frac{v^2}{r}$, where $v$ is the magnitude of the velocity, then a flat rotation curve for a galaxy can be reproduced only if $\nabla\phi\propto\frac{1}{r}$. But it seems that this case the behavior is not reproduced for any value of the parameter $C$. From eq. (\ref{eq:68}), it is clear that the Newtonian behavior is reproduced as $C=0$. For a well behaved solution, from the table \ref{tab2}, we can see that in such a case, $\omega=0$. From eq. (\ref{eq:60}), the relation $\frac{C}{\omega}$, then becomes:

\begin{equation}   \label{eq:70}
\frac{C}{\omega}=\frac{1}{\frac{29}{9}f_0-\frac{35}{18}-\frac{16}{9f_0}}=\frac{0}{0}
\end{equation}
 
Where we have introduced the appropriate value for $\gamma$ taken from the table \ref{tab2}. If we replace this condition inside eq. (\ref{eq:68}), we then obtain:

\begin{equation}   \label{eq:71}
\nabla\phi=-\left(\frac{A}{4}\right)\left(\frac{1}{\frac{29}{9}f_0-\frac{35}{18}-\frac{16}{9f_0}}\right)\left(\frac{1}{r^2}\right)=-\frac{GM}{r^2}
\end{equation}

Where we have imposed the Newtonian limit condition. We have to satisfy the condition:

\begin{equation}   \label{eq:72}
GM=\left(\frac{A}{4}\right)\left(\frac{1}{-\frac{29}{9}f_0+\frac{35}{18}+\frac{16}{9f_0}}\right)
\end{equation}

Then equation (\ref{eq:0}), for the full potential becomes:

\begin{equation}   \label{eq:73}
\phi=-\frac{GM}{r}+\frac{15f_0}{32-29f_0}
\end{equation}

Where we have replaced the appropriate values for the constant term in eq. (\ref{eq:0}). From the previous equations, it is clear that if we want to reproduce the appropriate Newtonian behavior, then the constant $A$ has to satisfy:

\begin{equation}   \label{eq:74}
A=4GM\left(-\frac{29}{9}f_0+\frac{35}{18}+\frac{16}{9f_0}\right)
\end{equation}

The remaining constant term is not important in order to obtain the Newtonian behavior. It is just a constant quantity which can be ignored for the computations. 

\subsection{The case $C=1$}

The case $C=1$ is extremely relevant since it looks like a Yukawa-like potential. If we replace the appropriate value for $\gamma$ and $\omega$ from the Table \ref{tab2}, then we can write the equation (\ref{eq:69}) like: 

\begin{equation}   \label{eq:75}
\nabla\phi=-\left(\frac{A}{4}\right)\frac{D}{\left(1-2f_0-\frac{3}{32}(23f_0-\sqrt{f_0}\sqrt{-160+209f_0}\right)}\frac{e^{-D/r}}{r^2}
\end{equation}

Where we have used the first value for $\omega$ corresponding to $C=1$. For the second value of $\omega$ corresponding to $C=1$, we can obtain the following result:

\begin{equation}   \label{eq:76}
\nabla\phi=-\left(\frac{A}{4}\right)\frac{D}{\left(1-2f_0-\frac{3}{32}(23f_0+\sqrt{f_0}\sqrt{-160+209f_0}\right)}\frac{e^{-D/r}}{r^2}
\end{equation}   

The form of this solution just suggest that the behavior of this potential is approximately Newtonian as the exponential factor tends to 1. The attractive or repulsive character of this solution depends on the values taken by $f_0$.

\subsection{The case $C=-1$}

The case $C=-1$ is perhaps the most interesting for our present purpose. This case is interesting since the field equation (\ref{eq:64}) becomes:

\begin{equation}   \label{eq:kiko}
\mu\nabla^2\phi=-\nabla\phi\cdot\nabla\mu
\end{equation}

This equation has can be written as:

\begin{equation}   \label{eq:kiko1}
\nabla\cdot(\mu\nabla\phi)=0
\end{equation}   
   
In vacuum, this has the same structure as the equation (\ref{eq:1006}). With the difference that in the non-local model the interpolating function $\mu$ is a function of the potential itself, rather than a function of its gradient as MOND suggest. The case $C=-1$ requires $B\neq0$ and $B>\frac{2}{r}$ in eq. \ref{eq:65}, otherwise the potential in such a case becomes complex.

\section{A comparison with other models}   \label{eq:comp}

The non-local model analyzed in this manuscript is able to reproduce the equation (\ref{eq:17}) with the definitions (\ref{eq:55}). This equation (after some arrangements) is similar to eq. (\ref{eq:1005}) or (\ref{eq:13}) which is obtained from the AQUAL Lagrangian (\ref{eq:1004}). However, the present model cannot reproduce the same dynamics due to MOND since the predicted interpolating function $\mu$ is a function of the potential ($\phi$), rather than a function of the acceleration ($\vert\nabla\phi\vert$) as can be observed from eq. (\ref{eq:55}) and the fact that $f(\psi)=f_0e^{\gamma\psi}$ with $\psi=-2\phi$. This previous relation is precisely the source of the problem for reproducing MOND appropriately. In \cite{M1}, it has been demonstrated that in order to reproduce the MONDian dynamics, it is necessary to add to the Einstein-Hilbert action, a Lagrangian such that it cancels the quadratic parts of the action and also provides some additional terms whose variations are:

\begin{equation}   \label{eq:a1}
\frac{c^2}{2a_0r^2}((rb'(r))^2)'=\frac{8\pi G\rho}{c^4}
\end{equation}
       
\begin{equation}   \label{eq:a2}
\frac{c^2}{a_0r^3}(krb'(r)-a(r))^2=0
\end{equation}

where $a(r)$ and $b(r)$ are given by:

\begin{equation}   \label{eq:a3}
a(r)\equiv A(r)-1\;\;\;\;\;b(r)\equiv B(r)-1
\end{equation}

with a static, spherically symmetric geometry defined by:

\begin{equation}   \label{eq:a4}
ds^2=-B(r)c^2dt^2+A(r)dr^2+r^2d\Omega^2
\end{equation}

the key point in the work performed in \cite{M1} is to change how the potentials depend upon the source without changing how they depend each other. In fact, the relation between the linearized potentials is given by:

\begin{equation}   \label{eq:a5}
a(r)\approx rb'(r)
\end{equation}

This relation is necessary in order to reproduce the appropriate amount of weak lensing consistent with the data. In the standard formalism of General Relativity, the linearized potentials take the form:

\begin{equation}   \label{eq:a6}
rb'(r)\approx \frac{2GM(r)}{c^2r}
\end{equation}

But in the MONDian regime, the following relation has to be satisfied:

\begin{equation}   \label{eq:a7}
rb'(r)\to\frac{2\sqrt{a_0 GM(r)}}{c^2}
\end{equation}

The MOND Lagrangian which cancels the quadratic terms of the Einstein-Hilbert action and also reproduces the results (\ref{eq:a1}) and (\ref{eq:a2}) is \cite{M1}:

\begin{multline}   \label{eq:a8}
L_{MOND}\to \frac{c^4}{16\pi G}\left(rab'(r)-\frac{a^2(r)}{2}+O(h^3)\right)\\
+\frac{c^2}{a_0}\left(\frac{\alpha a^3(r)}{r}+\beta a^2(r)b'(r)+\gamma ra(r)b'^2(r)+\delta r^2b'^3(r)+O(h^4)\right)
\end{multline}

the first line of this Lagrangian, just cancels the Einstein-Hilbert action terms given by:

\begin{equation}   \label{eq:a9}
L_{EH}=\frac{c^4}{16\pi G}R\sqrt{-g}\to \frac{c^4}{16\pi G}\left(-ra(r)b'(r)+\frac{a^2(r)}{2}+O(h^3)\right)
\end{equation}

Where the right-hand side (after the arrow) is obtained after integration by parts and here we ignore total derivative terms. Note that for the total Lagrangian $L=L_{EH}+L_{MOND}$, the Einstein-Hilbert terms vanishes. It has been demonstrated in \cite{M1} that no local invariant Lagrangian can reproduce the cubic terms of the MONDian action (\ref{eq:a8}). The reason is that the curvature tensor and its possible contractions, can only reproduce terms involving two derivatives acting on one or more weak fields in the following way \cite{M1}: 

\begin{equation}   \label{eq:a10}
(Curvature)^N\sim (h'')^N+O((h')^2(h'')^{N-1})
\end{equation}

On the other hand, the MOND corrections in eq. (\ref{eq:a8}), involve powers of just one derivative acting on a single weak field like:

\begin{equation}   \label{eq:a11}
L_{MOND}\sim\frac{c^4r^2}{16\pi G}\left((h')^2+\frac{c^2}{a_0}(h')^3+O(h^4)\right)
\end{equation}

is in this part where the model proposed in this manuscript fails. In this manuscript, the non-localities enter through the function $f(\psi)$ with the Lagrange multiplier condition $R=\square\psi$ (with $-2\phi=\psi$). In such a case, then the non-localities will enter as an algebraic expansion of the potentials. This can be seen in eq. (\ref{eq:55}) and the action (\ref{eq:2}) if we take into account that $f(\psi)=f_0e^{\gamma\psi}$. If we want to reproduce the MONDian dynamics, one possibility is for example to expand the function $f(\psi)$ around the scale defined as the geometric average of the Gravitational radius and the inverse of the acceleration scale $a_0$ characteristic of MOND, the scale is $r_0=\sqrt{\frac{GM}{a_0}}$. In such a case, we would get $f(\psi)\approx f_0+f_0\gamma(\nabla\psi)_{r=r_0}$. But doing this expansion just breaks the nature of the model since in such a case, we are imposing by hand the scale at which the MONDian regime applies rather than obtaining it. The model proposed here cannot reproduce the form for the Lagrangian $(\ref{eq:a8})$ or $(\ref{eq:a11})$ in a natural way. \\
In \cite{M1}, the non-localities are used in order to reduce the number of derivatives for the weak fields and particular components for the curvature are selected by using a time-like 4-vector obtained from the gradient of the invariant volume of the past light cone (see \cite{M1} for details). The reduction of the number of derivatives is the appropriate such that the MONDian Lagrangians (\ref{eq:a8}) or (\ref{eq:a11}) with only a single derivative of the weak fields can be reproduced. Remember that the standard Einstein-Hilbert Lagrangian can only reproduce powers of two derivatives acting on weak fields. In the model proposed in this manuscript, it seems that the reduction of derivatives is higher, such that instead of a MONDian action with single derivatives on weak fields, we have an action with no derivatives (only algebraic relations). However the model proposed in this manuscript does not have any problem with the lenses since the condition (\ref{eq:a5}) is satisfied.\\
Another attempt for reproducing MOND by using non-localities is done in \cite{M2}. In such a case, the model can reproduce galaxy rotation curves but not the observed gravitational lenses. The model in \cite{M2} proposes a Lagrangian given by:

\begin{equation}   \label{eq:a12}
L=\frac{c^4}{16\pi G}(R+c^{-4}a_0^2F(c^4a_0^{-2}g^{\mu \nu}\epsilon,_{\mu}\epsilon,_{\nu}))\sqrt{-g}
\end{equation} 

Where $\epsilon$ is the small potential as it is defined in \cite{M2}. The main point here is that an interpolating function $F(x)$ is introduced since the beginning and MOND is embedded inside this Lagrangian under the assumption that at small $x$ the MONDian dynamics appear. The factor inside the interpolating function is a kinetic term for the small potential and it makes easier to recover the MONDian dynamics.   
This is the main difference with respect to the model proposed in this manuscript where, as has been said before, the non-localities are introduced as algebraic expansion of the potential. If we want to mimic in some sense the model suggested in \cite{M2}, we would have to expand the function $f(\psi)$ around the neighborhood of some imposed scale as has been explained before in this section. The model in \cite{M2} also proposes the same relation (\ref{eq:a7}) but it cannot reproduce the appropriate lenses without the Dark Matter assumption. 

\section{Conclusions}

The non-local model in the present form can reproduce some additional attractive effects for some range of the parameter $\gamma$. The model cannot reproduce the MONDian dynamics without a strong tunning of the parameters. However, it can reproduce the AQUAL field equations with a scale-dependent interpolating function $\mu$ for some special case given by $C=-1$. That case however, requires the additional condition $B>\frac{2}{r}$ everywhere. The reproduction of the AQUAL equations is in agreement with Milgrom and Bekenstein, the first step for getting a Relativistic version of MOND. Every attempt in modifying gravity in order to reproduce the MONDian dynamics, must reproduce equations like the AQUAL Lagrangian explained before in this manuscript. Further research is needed in order to see whether or not is viable to reproduce the Dark Matter effects in agreement with non-localities. Another alternatives for the introduction of non-localities have been explored in \cite{M1} and \cite{M2}. In \cite{M1}, the non-localities were able to reproduce the MONDian dynamics and the appropriate gravitational lenses. In the case of \cite{M2}, the non-localities could reproduce the dynamics but not the lenses.\\\\

{\bf Acknowledgement}

The author would like to thank Jacob Bekenstein for a very useful correspondence as well as Misao Sasaki for useful discussions and comments. This work is supported by MEXT (The Ministry of Education, Culture, Sports, Science and Technology) in Japan and KEK Theory Center.

\end{document}